\documentclass[12pt]{article}
\usepackage{a4wide}
\usepackage{latexsym}
\usepackage{amsmath}
\usepackage{amsfonts}
\usepackage{amscd}
\usepackage{cite}

\usepackage{pslatex}
\usepackage{graphicx}
\usepackage[latin1]{inputenc}
\usepackage[T1]{fontenc}

\allowdisplaybreaks

\newcommand{\bq}{\begin{eqnarray}}
\newcommand{\eq}{\end{eqnarray}}
\newcommand{\eps}{\varepsilon}

\begin{document}

\thispagestyle{empty}

\begin{flushright}
  MITP/16-069
 \\ MaPhy-AvH/2016-14
\end{flushright}

\vspace{1.5cm}

\begin{center}
  {\Large\bf The kite integral to all orders in terms of elliptic polylogarithms\\
  }
  \vspace{1cm}
  {\large Luise Adams ${}^{a}$, Christian Bogner ${}^{b}$, Armin Schweitzer ${}^{b}$ and Stefan Weinzierl ${}^{a}$ \\
  \vspace{1cm}
      {\small ${}^{a}$ \em PRISMA Cluster of Excellence, Institut f{\"u}r Physik, }\\
      {\small \em Johannes Gutenberg-Universit{\"a}t Mainz,}\\
      {\small \em D - 55099 Mainz, Germany}\\
  \vspace{2mm}
      {\small ${}^{b}$ \em Institut f{\"u}r Physik, Humboldt-Universit{\"a}t zu Berlin,}\\
      {\small \em D - 10099 Berlin, Germany}\\
  } 
\end{center}

\vspace{2cm}

\begin{abstract}\noindent
  {
We show that the Laurent series of the two-loop kite integral in $D=4-2\eps$ space-time dimensions
can be expressed in each order of the series expansion
in terms of elliptic generalisations of (multiple) polylogarithms.
Using differential equations we present an iterative method to compute any desired order.
As an example, we give the first three orders explicitly. 
   }
\end{abstract}

\vspace*{\fill}

\newpage

\section{Introduction}
\label{sec:intro}

Precision calculations in particle physics require the computation of Feynman integrals.
There is a wide class of Feynman integrals, mainly related to massless theories, which can
be expressed in terms of multiple polylogarithms.
With the advance of computational techniques in recent years, these may be computed 
efficiently \cite{Moch:2001zr,Weinzierl:2002hv,Weinzierl:2004bn,Moch:2005uc,Bierenbaum:2003ud,Brown:2008,Panzer:2014caa,Bogner:2015unknown,Kotikov:1990kg,Kotikov:1991pm,Remiddi:1997ny,Gehrmann:1999as,Argeri:2007up,MullerStach:2012mp,Henn:2013pwa,Ablinger:2015tua}.
More challenging are Feynman integrals, which cannot be expressed in terms of multiple polylogarithms.
A prominent example is the two-loop sunrise integral with non-zero masses \cite{Broadhurst:1993mw,Berends:1993ee,Bauberger:1994nk,Bauberger:1994by,Bauberger:1994hx,Caffo:1998du,Laporta:2004rb,Groote:2005ay,Groote:2012pa,Bailey:2008ib,MullerStach:2011ru,Adams:2013nia,Bloch:2013tra,Adams:2014vja,Adams:2015gva,Adams:2015ydq,Remiddi:2013joa,Bloch:2016izu}.
Evaluating this integral one encounters elliptic generalisations of (multiple) polylogarithms
and every term of the Laurent expansion of the dimensionally regulated sunrise integral
may be expressed in terms of functions from this class \cite{Adams:2015ydq}.
It is therefore natural to ask, if further integrals may be computed systematically
within the class of elliptic generalisations of (multiple) polylogarithms
to all orders in the dimensional regularisation parameter.

The kite integral is a two-loop integral with five internal propagators, 
out of which three have a non-zero mass $m$,
while the other two propagators are massless.
This integral gives a two-loop contribution to the electron self-energy
and the corresponding Feynman graph is shown in fig.~(\ref{fig_kite_graph}).
The kite integral has non-trivial sub-topologies: Contracting the two massless propagators
one obtains the massive sunrise integral.
It is therefore expected, that the kite integral will involve elliptic generalisations
of (multiple) polylogarithms.
An integral representation of the kite integral has already been
considered by Sabry in 1962 \cite{Sabry:1962},
where it was noted that ``one cannot proceed any further with these integrals, as the modulus 
of the elliptic functions is a complicated expression ...''.
This paper is about ``yes we can'' proceed further.
Recently, Remiddi and Tancredi re-considered the kite integral
using dispersion relations \cite{Remiddi:2016gno}.
In particular, they present a method to re-write the ${\mathcal O}(\eps^j)$-term of the
Laurent expansion in $D=4-2\eps$ space-time dimensions as an iterated integral involving
complete elliptic integrals.
In this paper we show that the ${\mathcal O}(\eps^j)$-term of the
Laurent expansion can be expressed in terms of elliptic generalisations
of (multiple) polylogarithms and we present an algorithm to compute the 
${\mathcal O}(\eps^j)$-term.
In order to achieve this, we use differential equations, slightly modify the basis
of master integrals introduced in \cite{Remiddi:2016gno}, change the integration variable
from the momentum squared $t=p^2$ to the nome $q$ of the elliptic curve
and show that all integrations can be carried out within the class of
elliptic generalisations of polylogarithms.
The non-trivial part is the change of variables from $t$ to $q$.
A priori it is not obvious that after the change of variables 
all integration kernels are ``nice'' functions of the variable $q$.
We show that this is the case.

The structure of our result is also interesting in another aspect.
We recall that for the sunrise integral the ${\mathcal O}(\eps^j)$-term of the expansion around
$D=2-2\eps$ space-time dimensions is of the form
\bq
 \frac{\psi_1}{\pi} E^{(j)},
\eq
where $\psi_1$ is a solution of the homogeneous differential equation
and $E^{(j)}$ an expression in terms of elliptic generalisations of (multiple) polylogarithms.
The differential equation for the sunrise graph is of second order, reflecting 
the elliptic nature of this integral and $\psi_1$ is a period of the elliptic curve.

On the other hand, we find for the kite integral that the ${\mathcal O}(\eps^j)$-term
is an expression in terms of elliptic generalisations of (multiple) polylogarithms
without any prefactor $\psi_1/\pi$.
It should be noted that this distinction may seem artificial, as we may express in the equal mass
case $\psi_1$ in terms of elliptic polylogarithms.
It is however in line with the information from the differential equation:
The kite integral has a first-order differential equation.
With a suitable choice of a basis the differential equation may be put into a standard form
such that
\bq
 \frac{d}{dt} \mathrm{Kite}
 & = & 
 \eps \left( \frac{1}{t} - \frac{2}{t-m^2} \right) \mathrm{Kite}
 \;\; + \;\; \mbox{simpler topologies}.
\eq
The homogeneous solution of this differential equation at lowest order in $\eps$
is a constant and not $\psi_1$.

This paper is organised as follows:
In section~\ref{sec:master_integrals} we derive the differential equations for the master integrals
of the kite family.
Section~\ref{sec:def_elliptic} contains definitions related to the elliptic curve, 
relevant to the kite integral and to the sub-topologies of the sunrise type.
Elliptic generalisations of multiple polylogarithms are discussed in section~\ref{sec:elliptic_polylogs}.
In section~\ref{sec:boundary} we give the boundary values at $t=0$ for all master integrals.
An essential part of our method is to change the integration variable form $t$ to $q$. This is
discussed in section~\ref{sec:change_of_variable}.
In section~\ref{sec:integration_algo} we present the integration algorithm, which allows us to obtain iteratively
the ${\mathcal O}(\eps^j)$-term of the solution.
Explicit results up to ${\mathcal O}(\eps^2)$ are presented in section~\ref{sec:results}.
Finally, section~\ref{sec:conclusions} contains our conclusions.


\section{Master integrals and differential equation}
\label{sec:master_integrals}

In $D$-dimensional Minkowski space the integral family for the kite integral 
\begin{figure}
\begin{center}
\includegraphics[scale=1.0]{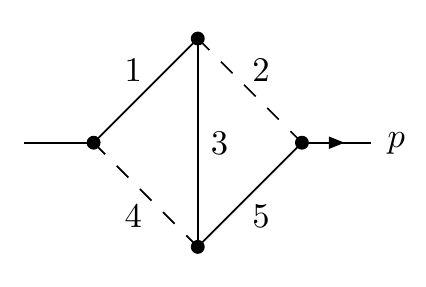}
\end{center}
\caption{
\it The kite graph. Solid lines correspond to massive propagators of mass $m$, dashed lines correspond to massless propagators.
}
\label{fig_kite_graph}
\end{figure}
is given by 
\bq
\label{def_kite}
 I_{\nu_1 \nu_2 \nu_3 \nu_4 \nu_5}\left( D, p^2, m^2, \mu^2 \right)
 & = &
 \left(-1\right)^{\nu_{12345}}
 \left(\mu^2\right)^{\nu_{12345}-D}
 \int \frac{d^Dk_1}{i \pi^{\frac{D}{2}}} \frac{d^Dk_2}{i \pi^{\frac{D}{2}}}
 \frac{1}{D_1^{\nu_1} D_2^{\nu_2} D_3^{\nu_3} D_4^{\nu_4} D_5^{\nu_5}},
\eq
with the propagators
\bq
 D_1=k_1^2-m^2, \hspace{0.3cm}  
 D_2=k_2^2, \hspace{0.3cm}  
 D_3 = (k_1-k_2)^2-m^2, \hspace{0.3cm} 
 D_4=(k_1-p)^2, \hspace{0.3cm}  
 D_5 = (k_2-p)^2-m^2
\eq
and $\nu_{12345}=\nu_1+\nu_2+\nu_3+\nu_4+\nu_5$.
The internal momenta are denoted by $k_1$ and $k_2$, 
the internal mass by $m$ and the external momentum by $p$. 
The kite graph is shown in fig.~(\ref{fig_kite_graph}).
Let us further define 
\bq
t=p^2.
\eq
For the convenience of the reader, 
we will suppress the dependence of the integrals on the mass $m$ and the scale $\mu$ in the following 
and we write instead
\bq
 I_{\nu_1 \nu_2 \nu_3 \nu_4 \nu_5}(D,t) 
 & = & 
 I_{\nu_1 \nu_2 \nu_3 \nu_4 \nu_5}(D,t,m^2,\mu^2).
\eq
There are eight master integrals. 
\begin{figure}
\begin{center}
\includegraphics[scale=1.0]{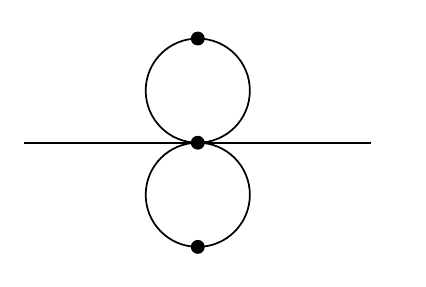}
\hspace*{5mm}
\includegraphics[scale=1.0]{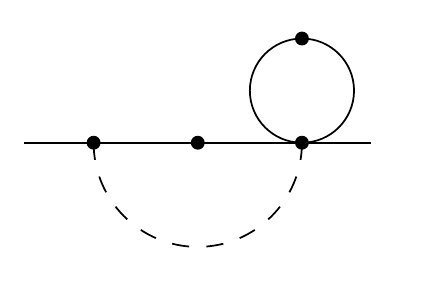}
\\
\includegraphics[scale=1.0]{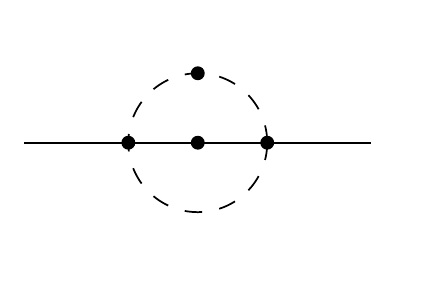}
\hspace*{5mm}
\includegraphics[scale=1.0]{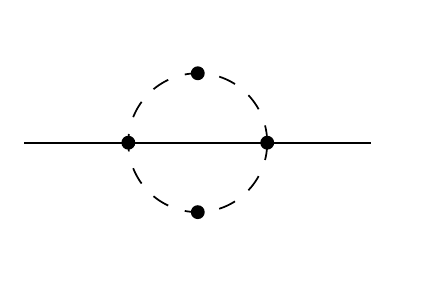}
\\
\includegraphics[scale=1.0]{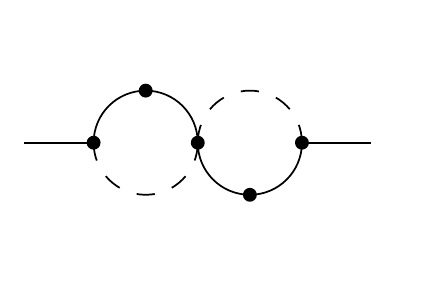}
\hspace*{5mm}
\includegraphics[scale=1.0]{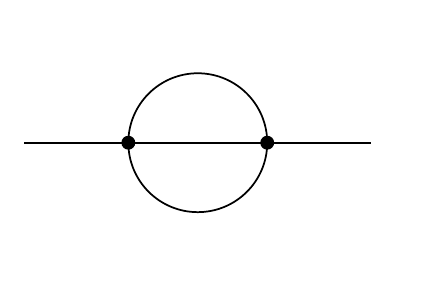}
\\
\includegraphics[scale=1.0]{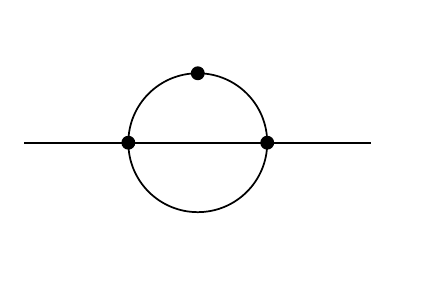}
\hspace*{5mm}
\includegraphics[scale=1.0]{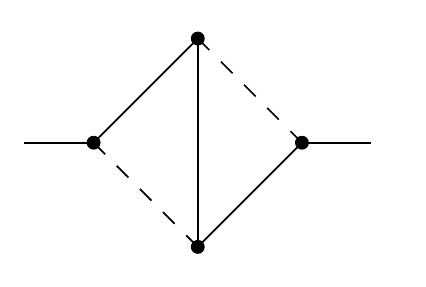}
\end{center}
\caption{
\it A set of master integrals for the kite family. 
A dot on a propagator indicates, that this propagator is raised to the power two.
The graphs correspond to the following integrals (from upper left to lower right):
$I_{2,0,2,0,0}$, $I_{2,0,2,1,0}$,
$I_{0,2,2,1,0}$, $I_{0,2,1,2,0}$,
$I_{2,1,0,1,2}$, $I_{1,0,1,0,1}$,
$I_{2,0,1,0,1}$ and $I_{1,1,1,1,1}$.
We use a linear combination of these master integrals as a basis.
}
\label{fig_master_integrals}
\end{figure}
A set of master integrals is shown in fig.~(\ref{fig_master_integrals}).
It will be convenient to use linear combinations of the integrals shown in fig.~(\ref{fig_master_integrals})
as a basis for the kite family in $D=4-2\eps$ space-time dimensions.
We will use the basis
\bq
\label{def_basis}
\lefteqn{
 I_1\left(D,t\right)
 =
 \left(D-4\right)^2 I_{20200}\left(D,t\right),
} & &
 \nonumber \\
\lefteqn{
 I_2\left(D,t\right)
 =
 \left(D-4\right)^2 \frac{t}{\mu^2} I_{20210}\left(D,t\right),
} & &
 \nonumber \\
\lefteqn{
 I_3\left(D,t\right)
 =
 \left(D-4\right)^2 \frac{t}{\mu^2} I_{02210}\left(D,t\right),
} & &
 \nonumber \\
\lefteqn{
 I_4\left(D,t\right)
 =
 \left(D-4\right)^2 \left[ 2 \frac{m^2}{\mu^2} I_{02210}\left(D,t\right) - \frac{t-m^2}{\mu^2} I_{02120}\left(D,t\right) \right],
} & &
 \nonumber \\
\lefteqn{
 I_5\left(D,t\right)
 =
 \left(D-4\right)^2 \left(\frac{t}{\mu^2}\right)^2 I_{21012}\left(D,t\right),
} & &
 \nonumber \\
\lefteqn{
 I_6\left(D,t\right)
 =
 \frac{3 \left(D-4\right)\left(D-5\right) \mu^2 t}{\left(t-m^2\right)\left(t-9m^2\right)}
 \left[ 
         \frac{\left(3m^2-t\right)}{\mu^2} I_{20200}\left(D,t\right)
 \right.
} & &
 \nonumber \\
 & &
 \left.
         + \left(3 D - 8 \right) \left(D-3\right) I_{10101}\left(D,t\right)
         + 2 \left(D-3\right) \frac{\left(t+3m^2\right)}{\mu^2} I_{20101}\left(D,t\right)
 \right],
 \nonumber \\
\lefteqn{
 I_7\left(D,t\right)
 =
 \frac{2 \left(D-4\right) \mu^6 t}{\left(t-m^2\right)^2\left(t-9m^2\right)^2}
} & &
 \nonumber \\
 & &
 \left\{
         \frac{\left(D-4\right)t^3 - \left(17D-71\right) m^2 t^2 + 3 \left(9D-46\right) m^4 t - 27 \left(D-5\right) m^6}{\mu^6} I_{20200}\left(D,t\right)
 \right.
 \nonumber \\
 & &
 \left.
         + \left(D-3\right) \left(3D-8\right) \frac{\left(D-3\right)t^2 - 10 m^2 t - 9 \left(D-5\right)m^4}{\mu^4} I_{10101}\left(D,t\right)
 \right.
 \nonumber \\
 & &
 \left.
         - \left(D-3\right) \frac{\left(D-4\right)t^3 -6 \left(6D-23\right)m^2 t^2 +15\left(3D-8\right) m^4 t + 54 \left(D-5\right) m^6}{\mu^6} I_{20101}\left(D,t\right)
 \right\},
 \nonumber \\
\lefteqn{
 I_8\left(D,t\right)
 =
 \left(D-4\right)\left(D-5\right) \frac{t}{\mu^2} I_{02210}\left(D,t\right)
 + \left(D-3\right)\left(D-4\right)^2 \left(D-5\right) \frac{t}{\mu^2} I_{11111}\left(D,t\right).
} & &
\eq
The basis integrals $I_6(D,t)$ and $I_7(D,t)$ are related to the two master integrals for the sunrise graph with three non-zero masses.
The particular linear combinations appearing in $I_6(D,t)$ and $I_7(D,t)$ can be expressed as sunrise integrals in $D-2=2-2\eps$ space-time dimensions.
We have
\bq
 I_6\left(D,t\right)
 & = &
 \left(D-4\right)\left(D-5\right) \frac{t}{\mu^2} I_{10101}\left(D-2,t\right),
 \nonumber \\
 I_7\left(D,t\right)
 & = &
 2 \left(D-4\right) \frac{m^2 t}{\mu^4} I_{20101}\left(D-2,t\right).
\eq
We denote the vector of basis integrals by $\vec{I}=(I_1,I_2,I_3,I_4,I_5,I_6,I_7,I_8)^T$.
With this choice of basis integrals the differential equation takes a particularly simple form \cite{Henn:2013pwa,Henn:2014qga}:
It is of Fuchsian type, where the only singularities are at $t\in\{0,m^2,9m^2\}$.
In addition, the right-hand side of the differential equation is linear in $\eps$.
The differential equation is easily obtained with the help of the program ``Reduze'' \cite{Studerus:2009ye,vonManteuffel:2012np} and reads
\bq
\label{dgl_t}
 \mu^2 \frac{d}{dt} \vec{I}
 & = &
 \left[ 
  \frac{\mu^2}{t} A_0 + \frac{\mu^2}{t-m^2} A_1 + \frac{\mu^2}{t-9m^2} A_9 \right] \vec{I},
\eq
where the $8 \times 8$-matrices $A_0$, $A_1$ and $A_9$ are linear in $\eps$.
They are given by
\bq
 A_0
 & = &
 \left(
 \begin{array}{rrrrrrrr}
 0 & 0 & 0 & 0 & 0 & 0 & 0 & 0 \\
 0 & \eps & 0 & 0 & 0 & 0 & 0 & 0 \\
 0 & 0 & \eps & 0 & 0 & 0 & 0 & 0 \\
 0 & 0 & -4 \eps & 0 & 0 & 0 & 0 & 0 \\
 0 & 0 & 0 & 0 & 2 \eps & 0 & 0 & 0 \\
 0 & 0 & 0 & 0 & 0 & -2 \eps & -\frac{3}{2}-3\eps & 0 \\
 0 & 0 & 0 & 0 & 0 & \frac{2}{3}+2\eps & 2+3\eps & 0 \\
 0 & 0 & 0 & 0 & -1-2\eps & -3\eps & 0 & \eps \\
 \end{array}
 \right),
 \nonumber \\
 A_1
 & = &
 \left(
 \begin{array}{rrrrrrrr}
 0 & 0 & 0 & 0 & 0 & 0 & 0 & 0 \\
 -\eps & -2\eps & 0 & 0 & 0 & 0 & 0 & 0 \\
 0 & 0 & -2 \eps & \eps & 0 & 0 & 0 & 0 \\
 0 & 0 & 4 \eps & -2 \eps & 0 & 0 & 0 & 0 \\
 0 & -2 \eps & 0 & 0 & -4 \eps & 0 & 0 & 0 \\
 0 & 0 & 0 & 0 & 0 & 0 & 0 & 0 \\
 -\frac{\eps}{4} & 0 & 0 & 0 & 0 & -\frac{1}{2}-\frac{3}{2}\eps & -1-2\eps & 0 \\
 \frac{1}{2}+\eps & 0 & -1-2\eps & 0 & 0 & \frac{8}{3}\eps & 0 & -2\eps \\
 \end{array}
 \right),
 \nonumber \\
 A_9
 & = &
 \left(
 \begin{array}{rrrrrrrr}
 0 & 0 & 0 & 0 & 0 & 0 & 0 & 0 \\
 0 & 0 & 0 & 0 & 0 & 0 & 0 & 0 \\
 0 & 0 & 0 & 0 & 0 & 0 & 0 & 0 \\
 0 & 0 & 0 & 0 & 0 & 0 & 0 & 0 \\
 0 & 0 & 0 & 0 & 0 & 0 & 0 & 0 \\
 0 & 0 & 0 & 0 & 0 & 0 & 0 & 0 \\
 \frac{9}{4} \eps & 0 & 0 & 0 & 0 & -\frac{1}{6}-\frac{\eps}{2} & -1-2\eps & 0 \\
 0 & 0 & 0 & 0 & 0 & 0 & 0 & 0 \\
 \end{array}
 \right).
\eq
Note that all matrices have a lower triangle block form.
The basis integrals $\{I_1,I_2,I_3,I_4,I_5\}$ form a closed sub-system. In this sub-system the right-hand side of the differential
equation is proportional to $\eps$.
This sub-system can be solved in terms of multiple polylogarithms.
In principle, we may further decompose the system of the first five basis integrals into smaller sub-systems.
However, the system for the first five basis integrals is already in a form, which allows an easy solution and there is no need
for any further refinements. 
For completeness we note that the set $\{I_3,I_4\}$ forms a closed sub-system, and so do the sets $\{I_1\}$, $\{I_1,I_2\}$ and $\{I_1,I_2,I_5\}$.

The basis integrals $\{I_1,I_6,I_7\}$ form another closed sub-system. These basis integrals are
the three basis integrals for the two-loop sunrise family with three massive propagators.
In this sub-system the right-hand side of the differential equation contains an $\eps^0$-term, 
which cannot be removed by a change of basis, indicating that
the result for $I_6$ and $I_7$ cannot be expressed in terms of multiple polylogarithms.
Here, elliptic generalisations of multiple polylogarithms occur \cite{MullerStach:2011ru,Adams:2013nia,Bloch:2013tra,Adams:2014vja,Adams:2015gva,Adams:2015ydq}.

Finally, considering $I_8$ requires the full system of lower-point integrals.

It is easily checked that the $\eps$-expansion of the basis integrals $\{I_1,I_2,I_3,I_4,I_5,I_6,I_7,I_8\}$ starts at $\eps^0$ at the earliest.
The basis integrals $I_1$, $I_4$ and $I_8$ start at $\eps^0$, the basis integrals $I_2$, $I_3$, $I_6$ and $I_7$ start at $\eps^1$, while
the basis integral $I_5$ starts at $\eps^2$.
For all basis integrals we write
\bq
 I_k\left(4-2\eps,t\right)
 & = &
 e^{-2\gamma_E \eps} \sum\limits_{j=0}^\infty \eps^j  I_k^{(j)}\left(4,t\right),
 \;\;\;\;\;\;
 k = 1, ..., 8,
\eq
where $\gamma_E$ is Euler's constant.


\section{Definitions related to the elliptic curve}
\label{sec:def_elliptic}

The differential equation for the kite integral contains the sunrise integral.
The latter is associated to an elliptic curve.
In Feynman parameter space, this elliptic curve is given by the cubic equation
\bq
\label{def_elliptic_curve}
 - x_1 x_3 x_5 t + m^2 \left( x_1 + x_3 + x_5 \right) \left( x_1 x_3 + x_3 x_5 + x_5 x_1 \right)
 & = & 0,
\eq
together with the choice of a rational point as origin.
The cubic equation in eq.~(\ref{def_elliptic_curve}) can be transformed into the Weierstrass normal form
\bq
 y^2 & = & 4 \left(x-e_1\right) \left(x-e_2\right) \left(x-e_3\right),
\eq
where the roots are given by
\bq
\label{def_roots}
 e_1 
 & = & 
 \frac{1}{24 \mu^4} \left( -t^2 + 6 m^2 t + 3 m^4 + 3 \sqrt{\tilde{D}} \right),
 \nonumber \\
 e_2 
 & = & 
 \frac{1}{24 \mu^4} \left( -t^2 + 6 m^2 t + 3 m^4 - 3 \sqrt{\tilde{D}} \right),
 \nonumber \\
 e_3 
 & = & 
 \frac{1}{24 \mu^4} \left( 2 t^2 - 12 m^2 t - 6 m^4 \right),
\eq
with
\bq
 \tilde{D} & = & \left( t - m^2 \right)^3 \left( t - 9 m^2 \right).
\eq
The modulus $k$ and the complementary modulus $k'$ of the elliptic curve are
\bq
 k = \sqrt{\frac{e_3-e_2}{e_1-e_2}},
 & &
 k' = \sqrt{1-k^2} = \sqrt{\frac{e_1-e_3}{e_1-e_2}}.
\eq
The periods of the elliptic curve related to the holomorphic one-form $dx/y$ are
\bq
\label{def_periods}
 \psi_1 =  
 2 \int\limits_{e_2}^{e_3} \frac{dx}{y}
 =
 \frac{4 \mu^2}{\tilde{D}^{\frac{1}{4}}} K\left(k\right),
 & &
 \psi_2 =  
 2 \int\limits_{e_1}^{e_3} \frac{dx}{y}
 =
 \frac{4 i \mu^2}{\tilde{D}^{\frac{1}{4}}} K\left(k'\right),
\eq
where $K(x)$ denotes the complete elliptic integral of the first kind
\bq
 K(x)
 & = &
 \int\limits_0^1 \frac{dt}{\sqrt{\left(1-t^2\right)\left(1-x^2t^2\right)}}.
\eq
The nome is defined by
\bq
 q \;\; = \;\; e^{i\pi \tau},
 & \mbox{with} &
 \tau \;\; = \;\; \frac{\psi_2}{\psi_1}.
\eq
The Wronskian is given by
\bq
 W & = &
 \psi_1 \frac{d}{dt} \psi_2 - \psi_2 \frac{d}{dt} \psi_1
 =
 -
 \frac{12 \pi i \mu^4}{t\left( t - m^2 \right)\left( t - 9 m^2 \right)}.
\eq
We further denote by $r_n$ the $n$-th root of unity
\bq
 r_n & = & e^{\frac{2\pi i}{n}}.
\eq
In particular we will need the third root of unity
\bq
 r_3 & = & e^{\frac{2 \pi i}{3}} 
 \;\; = \;\;
 \frac{1+i\sqrt{3}}{1-i\sqrt{3}}
 \;\; = \;\;
 - \frac{1}{2} + \frac{i}{2} \sqrt{3}.
\eq


\section{Elliptic generalisations of polylogarithms}
\label{sec:elliptic_polylogs}

In this section we define elliptic generalisations of (multiple) polylogarithms.
These generalisations appeared already in the all-order result for the sunrise integral \cite{Adams:2015ydq}.

The first five basis integrals can be expressed in terms of multiple polylogarithms.
Therefore it is useful to review these first.
Multiple polylogarithms are defined by
\bq
\label{def_multiple_polylogs}
 \mathrm{Li}_{n_1,n_2,...,n_l}\left(x_1,x_2,...,x_l\right)
 & = &
 \sum\limits_{j_1=1}^\infty \sum\limits_{j_2=1}^{j_1-1} ... \sum\limits_{j_l=1}^{j_{l-1}-1}
 \frac{x_1^{j_1}}{j_1^{n_1}} \frac{x_2^{j_2}}{j_2^{n_2}} ... \frac{x_l^{j_l}}{j_l^{n_l}}.
\eq
The sum representation is convergent for 
\bq
\label{condconvLi}
\left| x_1 x_2 ... x_j \right| \le 1 & & \mbox{for all} \; j \in \{1,...,l\} \; \mbox{and} \; (n_1,x_1) \neq(1,1).
\eq
In addition, there is an integral representation for multiple polylogarithms, given 
for $z_l \neq 0$ by:
\bq
\label{Gfuncdef}
G(z_1,...,z_l;y) & = &
 \int\limits_0^y \frac{dt_1}{t_1-z_1}
 \int\limits_0^{t_1} \frac{dt_2}{t_2-z_2} ...
 \int\limits_0^{t_{l-1}} \frac{dt_l}{t_l-z_l}.
\eq
With the notation
\bq
\label{Gshorthand}
G_{n_1,...,n_l}(z_1,...,z_l;y)
 & = &
 G(\underbrace{0,...,0}_{n_1-1},z_1,...,z_{l-1},\underbrace{0...,0}_{n_l-1},z_l;y),
\eq
where all $z_j$ for $j=1,...,l$ are assumed to be non-zero, the relation between the two notations reads
\bq
\label{Gintrepdef}
\mathrm{Li}_{n_1,...,n_l}(x_1,...,x_l)
& = & (-1)^l
 G_{n_1,...,n_l}\left( \frac{1}{x_1}, \frac{1}{x_1 x_2}, ..., \frac{1}{x_1...x_l};1 \right).
\eq
Let us now turn to elliptic generalisations of polylogarithms.
We define functions of $(2l+1)$ variables $x_1$, ..., $x_l$, $y_1$, ..., $y_l$, $q$
as follows:
For $l=1$ we set
\bq
 \mathrm{ELi}_{n;m}\left(x;y;q\right) & = & 
 \sum\limits_{j=1}^\infty \sum\limits_{k=1}^\infty \; \frac{x^j}{j^n} \frac{y^k}{k^m} q^{j k}.
\eq
For $l>1$ we define
\bq
\lefteqn{
 \mathrm{ELi}_{n_1,...,n_l;m_1,...,m_l;2o_1,...,2o_{l-1}}\left(x_1,...,x_l;y_1,...,y_l;q\right) 
 = }
 & & \nonumber \\
 & = &
 \sum\limits_{j_1=1}^\infty ... \sum\limits_{j_l=1}^\infty
 \sum\limits_{k_1=1}^\infty ... \sum\limits_{k_l=1}^\infty
 \;\;
 \frac{x_1^{j_1}}{j_1^{n_1}} ... \frac{x_l^{j_l}}{j_l^{n_l}}
 \;\;
 \frac{y_1^{k_1}}{k_1^{m_1}} ... \frac{y_l^{k_l}}{k_l^{m_l}}
 \;\;
 \frac{q^{j_1 k_1 + ... + j_l k_l}}{\prod\limits_{i=1}^{l-1} \left(j_i k_i + ... + j_l k_l \right)^{o_i}}.
\eq
We call these functions $\mathrm{ELi}$-functions.
We have the relations
\bq
\label{ELi_multiplication}
\lefteqn{
 \mathrm{ELi}_{n_1;m_1}\left(x_1;y_1;q\right) 
 \mathrm{ELi}_{n_2,...,n_l;m_2,...,m_l;2o_2,...,2o_{l-1}}\left(x_2,...,x_l;y_2,...,y_l;q\right) 
 = } & & \nonumber \\
 & &
 \hspace*{30mm}
 \mathrm{ELi}_{n_1,n_2,...,n_l;m_1,m_2,...,m_l;0,2o_2,...,2o_{l-1}}\left(x_1,x_2,...,x_l;y_1,y_2,...,y_l;q\right) 
 \hspace*{10mm}
\eq
and
\bq
\label{ELi_integration}
\lefteqn{
 \int\limits_0^q \frac{dq'}{q'}
 \mathrm{ELi}_{n_1,...,n_l;m_1,...,m_l;2o_1,2o_2,...,2o_{l-1}}\left(x_1,...,x_l;y_1,...,y_l;q'\right)
 = } & & \nonumber \\
 & &
 \hspace*{30mm}
 \mathrm{ELi}_{n_1,...,n_l;m_1,...,m_l;2(o_1+1),2o_2,...,2o_{l-1}}\left(x_1,...,x_l;y_1,...,y_l;q\right).
 \hspace*{10mm}
\eq
It will be convenient to introduce abbreviations for certain linear combinations, which occur quite often.
We define a prefactor $c_n$ and a sign $s_n$, both depending on an index $n$ by
\bq
 c_n = \frac{1}{2} \left[ \left(1+i\right) + \left(1-i\right)\left(-1\right)^n\right] = 
 \left\{ \begin{array}{rl}
 1, & \mbox{$n$ even}, \\
 i, & \mbox{$n$ odd}, \\
 \end{array} \right.
 & &
 s_n = (-1)^n =
 \left\{ \begin{array}{rl}
 1, & \mbox{$n$ even}, \\
 -1, & \mbox{$n$ odd}. \\
 \end{array} \right.
\eq
At depth $1$ we define the linear combinations
\bq
 \overline{\mathrm{E}}_{n;m}\left(x;y;q\right) 
 & = &
 \frac{c_{n+m}}{i}
 \left[
  \mathrm{ELi}_{n;m}\left(x;y;q\right)
  - s_{n+m} \mathrm{ELi}_{n;m}\left(x^{-1};y^{-1};q\right)
 \right].
\eq
More explicitly, we have
\bq
\label{def_Ebar_weight_1}
 \overline{\mathrm{E}}_{n;m}\left(x;y;q\right) 
 & = &
 \left\{ \begin{array}{ll}
 \frac{1}{i}
 \left[
 \mathrm{ELi}_{n;m}\left(x;y;q\right) - \mathrm{ELi}_{n;m}\left(x^{-1};y^{-1};q\right)
 \right],
 & \mbox{$n+m$ even,} \\
 & \\
 \mathrm{ELi}_{n;m}\left(x;y;q\right) + \mathrm{ELi}_{n;m}\left(x^{-1};y^{-1};q\right),
 & \mbox{$n+m$ odd.} \\
 \end{array}
 \right.
\eq
At higher depth we define functions
\bq
 \overline{\mathrm{E}}_{n_1,...,n_l;m_1,...,m_l;2o_1,...,2o_{l-1}}\left(x_1,...,x_l;y_1,...,y_l;q\right) 
\eq
as follows:
For $o_1=0$ we set
\bq
\label{Ebar_multiplication}
\lefteqn{
 \overline{\mathrm{E}}_{n_1,...,n_l;m_1,...,m_l;0,2o_2,...,2o_{l-1}}\left(x_1,...,x_l;y_1,...,y_l;q\right) 
 = } & & \nonumber \\
 & &
 \hspace*{30mm}
 \overline{\mathrm{E}}_{n_1;m_1}\left(x_1;y_1;q\right) 
 \overline{\mathrm{E}}_{n_2,...,n_l;m_2,...,m_l;2o_2,...,2o_{l-1}}\left(x_2,...,x_l;y_2,...,y_l;q\right).
 \hspace*{10mm}
\eq
For $o_1 > 0$ we set recursively
\bq
\label{Ebar_integration}
\lefteqn{
 \overline{\mathrm{E}}_{n_1,...,n_l;m_1,...,m_l;2(o_1+1),2o_2,...,2o_{l-1}}\left(x_1,...,x_l;y_1,...,y_l;q\right) 
 = 
 } & & \nonumber \\
 & &
 \hspace*{30mm}
 \int\limits_0^q \frac{dq'}{q'} 
 \overline{\mathrm{E}}_{n_1,...,n_l;m_1,...,m_l;2o_1,2o_2,...,2o_{l-1}}\left(x_1,...,x_l;y_1,...,y_l;q'\right).
 \hspace*{10mm}
\eq
Note that eq.~(\ref{Ebar_multiplication}) corresponds to eq.~(\ref{ELi_multiplication}),
and eq.~(\ref{Ebar_integration}) corresponds to eq.~(\ref{ELi_integration}).
The $\overline{\mathrm{E}}$-functions are linear combinations of the $\mathrm{ELi}$-functions
with the same indices.
More concretely, an $\overline{\mathrm{E}}$-function of depth $l$ can be expressed as a
linear combination of $2^l$ $\mathrm{ELi}$-functions.
We have
\bq
\lefteqn{
 \overline{\mathrm{E}}_{n_1,...,n_l;m_1,...,m_l;2o_1,...,2o_{l-1}}\left(x_1,...,x_l;y_1,...,y_l;q\right) 
 = 
 } & & \\
 & &
 \sum\limits_{t_1=0}^1 ... \sum\limits_{t_l=0}^1
 \left[ \prod\limits_{j=1}^l \frac{c_{n_j+m_j}}{i} \left( - s_{n_j+m_j} \right)^{t_j} \right]
 \mathrm{ELi}_{n_1,...,n_l;m_1,...,m_l;2o_1,...,2o_{l-1}}\left(x_1^{s_{t_1}},...,x_l^{s_{t_l}};y_1^{s_{t_1}},...,y_l^{s_{t_l}};q\right).
 \nonumber
\eq
Let us illustrate this with an example:
\bq
\lefteqn{
 \overline{\mathrm{E}}_{0,1;-1,0;2}\left(x_1,x_2;y_1,y_2;q\right)
 = 
 \mathrm{ELi}_{0,1;-1,0;2}\left(x_1,x_2;y_1,y_2;q\right)
 +
 \mathrm{ELi}_{0,1;-1,0;2}\left(x_1,x_2^{-1};y_1,y_2^{-1};q\right)
 } & &
 \nonumber \\
 & &
 \hspace*{27mm}
 +
 \mathrm{ELi}_{0,1;-1,0;2}\left(x_1^{-1},x_2;y_1^{-1},y_2;q\right)
 +
 \mathrm{ELi}_{0,1;-1,0;2}\left(x_1^{-1},x_2^{-1};y_1^{-1},y_2^{-1};q\right)
 \nonumber
\eq
There is a close relation between the $\overline{\mathrm{E}}$-functions
and the $\mathrm{E}$-functions introduced in \cite{Adams:2015ydq}, the difference being that the 
$\mathrm{E}$-functions at depth $1$ have a term proportional to $q^0$, while the
$\overline{\mathrm{E}}$ do not.
We have the relation
\bq
\label{relation_E_Ebar}
\lefteqn{
 \mathrm{E}_{n_1,...,n_{l-1},n_l;m_1,...,m_{l-1},m_l;2o_1,...,2o_{l-2},2o_{l-1}}\left(x_1,...,x_{l-1},x_l;y_1,...,y_{l-1},y_l;q\right)
  =
} & & \nonumber \\
 & &
\hspace*{15mm}
 \overline{\mathrm{E}}_{n_1,...,n_{l-1},n_l;m_1,...,m_{l-1},m_l;2o_1,...,2o_{l-2},2o_{l-1}}\left(x_1,...,x_{l-1},x_l;y_1,...,y_{l-1},y_l;q\right)
 \nonumber \\
 & &
\hspace*{15mm}
 +
 \overline{\mathrm{E}}_{n_1,...,n_{l-2},n_{l-1}+o_{l-1};m_1,...,m_{l-2},m_{l-1}+o_{l-1};2o_1,...,2o_{l-2}}\left(x_1,...,x_{l-1};y_1,...,y_{l-1};q\right)
 \nonumber \\
 & &
\hspace*{15mm}
 \times
 \frac{c_{n_l+m_l}}{2i} \left[ \mathrm{Li}_{n_l}\left( x_l \right) - s_{n_l+m_l} \mathrm{Li}_{n_l}\left( x_l^{-1} \right) 
 \right].
\eq


\section{Boundary values}
\label{sec:boundary}

We use the point $t=0$ to obtain boundary values for the differential equation.
Most of the basis integrals vanish at $t=0$ (in $D$ dimensions) due to an explicit prefactor of $t$.
We have
\bq
 I_2\left(D,0\right) 
 \;\; = \;\;
 I_3\left(D,0\right) 
 \;\; = \;\;
 I_5\left(D,0\right) 
 \;\; = \;\;
 I_6\left(D,0\right) 
 \;\; = \;\;
 I_7\left(D,0\right) 
 \;\; = \;\;
 I_8\left(D,0\right) 
 \;\; = \;\;
 0.
\eq
Only the integrals $I_1$ and $I_4$ have non-vanishing boundary values at $t=0$.
We find
\bq
 I_1\left(4-2\eps,0\right) 
 & = &
 4 \left( \frac{m^2}{\mu^2} \right)^{-2\eps} \Gamma\left(1+\eps\right)^2 ,
 \nonumber \\
 I_4\left(4-2\eps,0\right) 
 & = &
 4 \left( \frac{m^2}{\mu^2} \right)^{-2\eps} \Gamma\left(1+\eps\right) \Gamma\left(1+2\eps\right) \Gamma\left(1-\eps\right).
\eq
Setting $L=\ln(m^2/\mu^2)$ we have for the first few terms
\bq
 I_1\left(4-2\eps,0\right) 
 & = &
 e^{-2\gamma_E \eps}
 \left[
        4 - 8 L \eps + \left( \frac{2}{3} \pi^2 + 8 L^2 \right) \eps^2
 \right]
 + {\mathcal O}\left(\eps^3\right),
 \nonumber \\
 I_4\left(4-2\eps,0\right) 
 & = &
 e^{-2\gamma_E \eps}
 \left[
        4 - 8 L \eps + \left( 2 \pi^2 + 8 L^2 \right) \eps^2
 \right]
 + {\mathcal O}\left(\eps^3\right).
\eq


\section{Changing the kinematic variable $t$ to the nome $q$}
\label{sec:change_of_variable}

In order to obtain an iterative all-order solution for the kite integral we change the variable from $t=p^2$ to the nome $q$.
We have
\bq
 \mu^2 \frac{d}{dt}
 & = & \mu^2 \frac{i \pi W}{\psi_1^2} q \frac{d}{dq}.
\eq
The differential equation in eq.~(\ref{dgl_t}) translates then into a 
differential equation for $\vec{I}$ with respect to the variable $q$.
We obtain
\bq
\label{dgl_q}
 q \frac{d}{dq} \vec{I}
 & = &
 \left(
  \frac{1}{i \pi \mu^2} \frac{\psi_1^2}{W} \frac{\mu^2}{t} A_0
  +
  \frac{1}{i \pi \mu^2} \frac{\psi_1^2}{W} \frac{\mu^2}{t-m^2} A_1
  +
  \frac{1}{i \pi \mu^2} \frac{\psi_1^2}{W} \frac{\mu^2}{t-9m^2} A_9
 \right) \vec{I}.
\eq
All terms appearing on the right-hand side of eq.~(\ref{dgl_q}) can be expressed
as functions $\mathrm{ELi}_{0;-1}$ of the variable $q$. 
Explicitly we have
\bq
\label{integration_kernel_1}
 \frac{1}{i \pi} \frac{\psi_1^2}{W} \frac{1}{t}
 & = &
 1 
 - 4 \overline{\mathrm{E}}_{0;-1}\left(r_3;-1;-q\right),
 \\
 \frac{1}{i \pi} \frac{\psi_1^2}{W} \frac{1}{t-m^2}
 & = &
 - \frac{3}{2} \overline{\mathrm{E}}_{0;-1}\left(r_3;-1;-q\right)
 + \frac{3}{2} \overline{\mathrm{E}}_{0;-1}\left(r_3;1;-q\right)
 + 3 \overline{\mathrm{E}}_{0;-1}\left(-1;1;-q\right),
 \nonumber \\
 \frac{1}{i \pi} \frac{\psi_1^2}{W} \frac{1}{t-9m^2}
 & = &
  \frac{1}{2} \overline{\mathrm{E}}_{0;-1}\left(r_3;-1;-q\right)
 - \frac{9}{2} \overline{\mathrm{E}}_{0;-1}\left(r_3;1;-q\right)
 + 3 \overline{\mathrm{E}}_{0;-1}\left(-1;1;-q\right).
 \nonumber
\eq
The expression on the second line can be written in a slightly more compact form
\bq
\label{integration_kernel_1a}
 \frac{1}{i \pi} \frac{\psi_1^2}{W} \frac{1}{t-m^2}
 & = &
 3 \overline{\mathrm{E}}_{0;-1}\left(-1;1;-q\right)
 - 3 \overline{\mathrm{E}}_{0;-1}\left(r_6;1;-q\right),
\eq
at the expense of introducing the sixth root of unity.
These formulae can be used to express the basis integrals $I_1$, $I_2$, $I_3$, $I_4$ and $I_5$
in terms of $\overline{\mathrm{E}}$-functions (or $\mathrm{ELi}$-functions).
The basis integrals $I_6$ and $I_7$ are related to the sunrise graph with three massive propagators.
We will only need $I_6$. 
We note that
\bq
 I_6\left(4-2\eps,t\right)
 & = &
 \left(2\eps\right)\left(1+2\eps\right) \frac{t}{\mu^2} I_{10101}\left(2-2\eps,t\right)
\eq
and $I_{10101}(2-2\eps,t)$ can be written as 
\bq
 I_{10101}\left(2-2\eps,t\right)
 \;\; = \;\;
 S_{111}\left(2-2\eps,t\right)
 \;\; = \;\;
 \frac{\psi_1}{\pi} E_{111}\left(2-2\eps,q\right),
\eq
where $E_{111}\left(2-2\eps,q\right)$ can be expressed in terms of the $\mathrm{ELi}$-functions.
In ref.~\cite{Adams:2015ydq} an algorithm was presented to compute the expansion 
of $E_{111}\left(2-2\eps,q\right)$ to all orders in $\eps$ in terms of the functions $\mathrm{ELi}$
and we may take the results from there.
The basis integral $I_6$ enters the differential equation for the kite integral $I_8$
in the terms proportional to $A_0$ and $A_1$.
In order to remain within the class of functions $\mathrm{ELi}$ we factor off
\bq 
\label{prefactor_sunrise}
 \frac{t}{\mu^2} \frac{\psi_1}{\pi}
\eq
from $I_6$.
Thus we have to consider the two additional integration kernels
\bq
\label{integration_kernel_2}
 \frac{1}{i \pi \mu^2} \frac{\psi_1^2}{W} \frac{\psi_1}{\pi}
 & = &
  -6 \overline{\mathrm{E}}_{0;-2}\left(r_3;-1;-q\right),
 \nonumber \\
 \frac{1}{i \pi \mu^2} \frac{\psi_1^2}{W} \frac{\psi_1}{\pi} \frac{t}{t-m^2}
 & = &
  - \frac{27}{4} \overline{\mathrm{E}}_{0;-2}\left(r_3;-1;-q\right)
  - \frac{27}{4} \overline{\mathrm{E}}_{0;-2}\left(r_3;1;-q\right).
\eq
Again, we see that these integration kernels can be expressed in terms of the functions $\mathrm{ELi}$.

For completeness we note that $\psi_1/\pi$ can be expressed in terms of $\mathrm{ELi}$-functions:
\bq
\label{psi_1_ELi_representation}
 \frac{\psi_1}{\pi}
 & = &
 \frac{2 \mu^2}{\sqrt{3} m^2}
 \left[ 1 
  + \frac{\sqrt{3}}{2} \overline{\mathrm{E}}_{0;0}\left(r_3;-1;-q\right)
  + \frac{3 \sqrt{3}}{2} \overline{\mathrm{E}}_{0;0}\left(r_3;1;-q\right)
 \right],
\eq
although we will only use eq.~(\ref{integration_kernel_1}) and eq.~(\ref{integration_kernel_2}), 
but not eq.~(\ref{psi_1_ELi_representation}).


\section{The integration algorithm}
\label{sec:integration_algo}

Combining eq.~(\ref{ELi_multiplication}) and eq.~(\ref{ELi_integration})
we have
\bq
\label{ELi_iteration}
\lefteqn{
 \int\limits_0^q \frac{dq'}{q'}
 \mathrm{ELi}_{n_1;m_1}\left(x_1;y_1;q'\right) 
 \mathrm{ELi}_{n_2,...,n_l;m_2,...,m_l;2o_2,...,2o_{l-1}}\left(x_2,...,x_l;y_2,...,y_l;q'\right) 
 = } & & \nonumber \\
 & &
 \hspace*{30mm}
 \mathrm{ELi}_{n_1,n_2,...,n_l;m_1,m_2,...,m_l;2,2o_2,...,2o_{l-1}}\left(x_1,x_2,...,x_l;y_1,y_2,...,y_l;q\right).
 \hspace*{10mm}
\eq
Eq.~(\ref{ELi_iteration}) is the essential equation, which allows us to integrate the differential equation 
to all orders in $\eps$.
Eq.~(\ref{ELi_iteration}) shows that as long as all integration kernels are of the form $\mathrm{ELi}_{n;m}(x;y;q)$, 
we will not leave the class of $\mathrm{ELi}$-functions.
In practice it is more convenient to work with the $\overline{\mathrm{E}}$-functions.
Taking into account that the argument of the $\overline{\mathrm{E}}$-functions is $(-q)$ instead of $q$,
eq.~(\ref{ELi_iteration}) translates to
\bq
\label{Ebar_iteration}
\lefteqn{
 \int\limits_0^{q} \frac{dq'}{q'}
 \overline{\mathrm{E}}_{n_1;m_1}\left(x_1;y_1;-q'\right) 
 \overline{\mathrm{E}}_{n_2,...,n_l;m_2,...,m_l;2o_2,...,2o_{l-1}}\left(x_2,...,x_l;y_2,...,y_l;-q'\right) 
 = } & & \nonumber \\
 & &
 \hspace*{30mm}
 \overline{\mathrm{E}}_{n_1,n_2,...,n_l;m_1,m_2,...,m_l;2,2o_2,...,2o_{l-1}}\left(x_1,x_2,...,x_l;y_1,y_2,...,y_l;-q\right).
 \hspace*{10mm}
\eq
In addition, we have the trivial integration rule
\bq
\label{Ebar_iteration2}
 \int\limits_0^{q} \frac{dq'}{q'}
 \overline{\mathrm{E}}_{n_1;m_1}\left(x_1;y_1;-q'\right) 
 & = &
 \overline{\mathrm{E}}_{n_1+1;m_1+1}\left(x_1;y_1;-q'\right).
\eq
Eq.~(\ref{Ebar_iteration}) and eq.~(\ref{Ebar_iteration2})
is all what is needed to integrate the differential equation.
In order to obtain the $\eps^j$-term of the kite integral, one proceeds through the following three
steps:
\begin{enumerate}
\item One first computes the first five basis integrals $I_1$ to $I_5$ up to order $\eps^j$
in terms of $\overline{\mathrm{E}}$-functions.
The basis integrals $\{I_1,I_2,I_3,I_4,I_5\}$ form a closed sub-system. 
The original differential equation~(\ref{dgl_t}) with respect to the variable $t$ 
is easily integrated in terms of harmonic polylogarithms.
However, for the kite integral we need these basis integrals in terms of 
the $\overline{\mathrm{E}}$-functions.
Given that we may express all integration kernels with the help of eq.~(\ref{integration_kernel_1}) and eq.~(\ref{integration_kernel_1a}) in terms of $\overline{\mathrm{E}}$-functions and using the integration rules
in eq.~(\ref{Ebar_iteration}) and eq.~(\ref{Ebar_iteration2}), the integration in the variable $q$ 
is as easy as the integration in the variable $t$.
\item In the second step we determine the function $E_{111}(2-2\eps,q)$, appearing in the basis integral
$I_6$ 
\bq
\label{I6_E111}
 I_6\left(4-2\eps,t\right)
 & = & 2 \eps \left(1+2\eps\right) \frac{t}{\mu^2} \frac{\psi_1}{\pi} E_{111}\left(2-2\eps,q\right)
\eq
to order $\eps^{j-2}$. Due to the prefactor $\eps$ in eq.~(\ref{I6_E111})
and another prefactor $\eps$ on the right-hand side of the differential equation 
we will need the expansion of $E_{111}(2-2\eps,q)$ only to order $\eps^{j-2}$.
The function $E_{111}(2-2\eps,q)$ is related to the sunrise integral around two space-time dimensions, after
the homogeneous solution $\psi_1$ has been factored out:
\bq
 S_{111}\left(2-2\eps,t\right) & = & 
 \frac{\psi_1}{\pi} E_{111}\left(2-2\eps,q\right).
\eq
The function $E_{111}\left(2-2\eps,q\right)$ has a Taylor expansion in $\eps$ 
\bq
 E_{111}\left(2-2\eps,q\right)
 & = &
 e^{-2\gamma_E\eps} \sum\limits_{j=0}^\infty \eps^j E_{111}^{(j)}(2,q),
\eq
and each term of this
Taylor expansion can be expressed in terms of the $\overline{\mathrm{E}}$-functions.
An algorithm to construct the $\eps^j$-term of $E_{111}(2-2\eps,q)$ has been given in \cite{Adams:2015ydq}.
What is important here is that every term of $E_{111}(2-2\eps,q)$ can be expressed in terms
of the $\overline{\mathrm{E}}$-functions.
\item Finally, we integrate the differential equation for $I_8$.
Assuming that we already computed all terms of order $\eps^k$ with $k<j$ for $I_8$, the differential
equation for the term of order $\eps^j$ of $I_8$ reads
\bq
\lefteqn{
 q \frac{d}{dq} I_8^{(j)}
 =  
 \left[  1 - 4 \overline{\mathrm{E}}_{0;-1}\left(r_3;-1;-q\right) \right]
 \left( -2 I_5^{(j-1)} - I_5^{(j)} + I_8^{(j-1)} \right)
 } & &
 \nonumber \\
 & &
 +
 3 \left[ \overline{\mathrm{E}}_{0;-1}\left(-1;1;-q\right) - \overline{\mathrm{E}}_{0;-1}\left(r_6;1;-q\right) \right]
 \left( I_1^{(j-1)} + \frac{1}{2} I_1^{(j)} -2 I_3^{(j-1)} - I_3^{(j)} - 2 I_8^{(j-1)} \right)
 \nonumber \\
 & &
  - 36 \overline{\mathrm{E}}_{0;-2}\left(r_3;1;-q\right) \left( E_{111}^{(j-2)} + 2 E_{111}^{(j-3)} \right).
\eq
Again, this integration can always be performed within the class of 
$\overline{\mathrm{E}}$-functions.
Note that the basis integral $I_7$ does not appear in the differential equation for $I_8$.
\end{enumerate}
We would like to add one remark:
Eq.~(\ref{integration_kernel_1}) and eq.~(\ref{integration_kernel_1a}) 
in combination with eq.~(\ref{Ebar_iteration}) and eq.~(\ref{Ebar_iteration2})
allow us to express any harmonic polylogarithm in the letters $\{0,1\}$ without trailing zeros
in terms of $\overline{\mathrm{E}}$-functions.
The converse is in general not true.


\section{Explicit results}
\label{sec:results}

In this section we present the explicit results for the first three terms in the $\eps$-expansion
for the basis integrals $I_1$-$I_5$, $I_6$ and $I_8$.

\subsection{The first five basis integrals}

The first five basis integrals can be expressed to all orders either in terms of multiple polylogarithms
or in terms of $\mathrm{ELi}$-functions.
We give here the first three terms of the $\eps$-expansion in both representations.
In terms of multiple polylogarithms we have with $L=\ln(m^2/\mu^2)$ and $y=t/m^2$
\bq
 I_1\left(4-2\eps,t\right)
 & = &
 e^{-2\gamma_E\eps} \left\{
        4 - 8 L \eps + \left( \frac{2}{3} \pi^2 + 8 L^2 \right) \eps^2
 \right\}
 + {\mathcal O}\left(\eps^3\right),
 \nonumber \\
 I_2\left(4-2\eps,t\right)
 & = &
 e^{-2\gamma_E\eps} \left\{
   - 4 G\left(1;y\right) \eps
   - 4 \left[ G\left(0,1;y\right) - 2 G\left(1,1;y\right) - 2 G\left(1;y\right) L \right] \eps^2
 \right\}
 + {\mathcal O}\left(\eps^3\right),
 \nonumber \\
 I_3\left(4-2\eps,t\right)
 & = &
 e^{-2\gamma_E\eps} \left\{
  4 G\left(1;y\right) \eps 
  + 4 \left[ G\left(0,1;y\right) - 4 G\left(1,1;y\right) - 2 G\left(1;y\right) L \right] \eps^2
 \right\}
 + {\mathcal O}\left(\eps^3\right),
 \nonumber \\
 I_4\left(4-2\eps,t\right)
 & = &
 e^{-2\gamma_E\eps} \left\{
  4
  - 8 \left[ G\left(1;y\right) + L \right] \eps
 \right. \nonumber \\
 & & \left.
  + 8 \left[ \frac{\pi^2}{4} - 2 G\left(0,1;y\right) + 4 G\left(1,1;y\right) + 2 G\left(1;y\right) L + L^2 \right] \eps^2
 \right\}
 + {\mathcal O}\left(\eps^3\right),
 \nonumber \\
 I_5\left(4-2\eps,t\right)
 & = &
 e^{-2\gamma_E\eps} \left\{
 8 G\left(1,1;y\right) \eps^2
 \right\}
 + {\mathcal O}\left(\eps^3\right).
\eq
The first five basis integrals depend up to the order $\eps^2$ only on the three
harmonic polylogarithms $G(1;y)$, $G(0,1;y)$ and $G(1,1;y)$.
In order to present the basis integrals $I_1$ to $I_5$ up to the order $\eps^2$ in terms of $\mathrm{ELi}$-functions 
it suffices to express these three harmonic polylogarithms in terms of $\mathrm{ELi}$-functions.
We have
\bq
\label{HPL_weight_1_2}
 G\left(1;y\right)
 & = &
 3 
 \left[ 
          \overline{\mathrm{E}}_{1;0}\left( -1;  1; -q \right)
        - \overline{\mathrm{E}}_{1;0}\left( r_6; 1; -q \right)
 \right],
 \nonumber \\
 G\left(0,1;y\right)
 & = &
 3 
 \left[ 
          \overline{\mathrm{E}}_{2;1}\left( -1;  1; -q \right)
        - \overline{\mathrm{E}}_{2;1}\left( r_6; 1; -q \right)
 \right]
 \nonumber \\
 & &
 - 12
 \left[ 
          \overline{\mathrm{E}}_{0,1;-1,0;2}\left( r_3,-1;  -1,1; -q \right)
        - \overline{\mathrm{E}}_{0,1;-1,0;2}\left( r_3,r_6; -1,1; -q \right)
 \right],
 \nonumber \\
 G\left(1,1;y\right)
 & = &
 9
 \left[ 
          \overline{\mathrm{E}}_{0,1;-1,0;2}\left( -1,-1;   1,1; -q \right)
        - \overline{\mathrm{E}}_{0,1;-1,0;2}\left( -1,r_6;  1,1; -q \right)
 \right.
 \nonumber \\
 & &
 \left.
        - \overline{\mathrm{E}}_{0,1;-1,0;2}\left( r_6,-1;  1,1; -q \right)
        + \overline{\mathrm{E}}_{0,1;-1,0;2}\left( r_6,r_6; 1,1; -q \right)
 \right].
\eq

\subsection{The sunrise integral}

With our particular choice of basis integrals in eq.~(\ref{def_basis}) only the first sunrise integral 
$I_6$, but not the second sunrise integral $I_7$ enters the differential equation for the kite integral.
The sunrise integral has been computed to all orders in $\eps$ in \cite{Adams:2015ydq} and we may take the result
for the first few terms from there.
We have
\bq
 I_6\left(4-2\eps,t\right)
 & = &
 \frac{t}{\mu^2} 
 \frac{\psi_1}{\pi} 
 e^{-2\gamma_E\eps} \left\{
 2 E_{111}^{(0)} \eps 
 + 2 \left[ E_{111}^{(1)} + 2 E_{111}^{(0)} \right] \eps^2
 \right\}
 + {\mathcal O}\left(\eps^3\right).
\eq
In accordance with the comment before eq.~(\ref{prefactor_sunrise}) we factor out a prefactor
$t \psi_1/(\mu^2 \pi)$.
$E^{(0)}_{111}$ and $E^{(1)}_{111}$ are given by
\bq
\lefteqn{
 E^{(0)}_{111}
 = 
 3 \mathrm{E}_{2;0}\left(r_3;-1;-q\right),
 } & &
 \nonumber \\
\lefteqn{
 E^{(1)}_{111}
 = 
 3 \mathrm{E}_{3;1}\left(r_3;-1;-q\right)
 + 3 \mathrm{E}_{0,1;-2,0;4}\left(r_3,r_3;-1,-1;-q\right)
 - 9 \mathrm{E}_{0,1;-2,0;4}\left(r_3,r_3;-1,1;-q\right)
 } & &
 \nonumber \\
 & &
 + 18 \mathrm{E}_{0,1;-2,0;4}\left(r_3,-1;-1,1;-q\right)
 + \frac{3}{2i} \left\{ 
                       - 2 \mathrm{Li}_{2,1}\left(r_3,1\right) - 2 \mathrm{Li}_3\left(r_3\right)
                       + 2 \mathrm{Li}_{2,1}\left(r_3^{-1},1\right) 
 \right. \nonumber \\
 & & \left.
                       + 2 \mathrm{Li}_3\left(r_3^{-1}\right)
                       + 6 \mathrm{Li}_1\left(-1\right) \left[ \mathrm{Li}_2\left(r_3\right) - \mathrm{Li}_2\left(r_3^{-1}\right) \right]
                \right\}
 + L_{1;0} E^{(0)}_{111},
\eq
with
\bq
 L_{1;0}
 & = &
         - 2 L
         - \mathrm{E}_{1;0}\left(r_3;-1;-q\right) 
         + 3 \mathrm{E}_{1;0}\left(r_3;1;-q\right) 
         - 6 \mathrm{E}_{1,0}\left(-1;1;-q\right).
\eq
The $\mathrm{E}$-functions have been defined in eq.~(\ref{relation_E_Ebar}).
The terms $E^{(j)}_{111}$ appear in the Taylor expansion of the sunrise integral around $D=2-2\eps$ space-time
dimensions:
\bq
 I_{10101}\left(2-2\eps,t\right)
 & = &
 e^{-2\gamma_E\eps} \frac{\psi_1}{\pi}
 \sum\limits_{j=0}^\infty E_{111}^{(j)} \eps^j.
\eq

\subsection{The kite integral}

The first three terms of the Taylor expansion of the basis integral $I_8$ read
\bq
\lefteqn{
 I_8\left(4-2\eps,t\right)
 =
 e^{-2\gamma_E\eps} \left\{
 2 G\left(1;y\right)
 + 
 2 \left[ G\left(0,1;y\right) - 4 G\left(1,1;y\right) - 2 G\left(1;y\right) L + 2 G\left(1;y\right) \right] \eps
 \right.
} & &
 \nonumber \\
 & &
 \left.
 +
 \left[
  32 G\left(1,1,1;y\right) 
  - 8 G\left(1,0,1;y\right)
  - 16 G\left(0,1,1;y\right) 
  + 2 G\left(0,0,1;y\right)
  - 16 G\left(1,1;y\right) 
 \right. \right.
 \nonumber \\
 & & \left. \left.
  + 4 G\left(0,1;y\right)
  + 16 G\left(1,1;y\right) L 
  - 4 G\left(0,1;y\right) L
  + \left( \frac{\pi^2}{3} - 8 L + 4 L^2 \right) G\left(1;y\right)
 \right. \right.
 \nonumber \\
 & & \left. \left.
 - 108 \mathrm{Cl}_2\left(\frac{2\pi}{3}\right) \overline{\mathrm{E}}_{1;-1}\left(r_3;1;-q\right)
 - 108 \overline{\mathrm{E}}_{0,2;-2,0;2}\left(r_3,r_3;1,-1;-q\right)
 \right] \eps^2
 \right\}
 + {\mathcal O}\left(\eps^3\right),
 \hspace*{5mm}
\eq
with the Clausen value
\bq
 \mathrm{Cl}_2\left(\frac{2\pi}{3}\right)
 & = &
 \frac{1}{2i} \left[ \mathrm{Li}_2\left(r_3\right) - \mathrm{Li}_2\left(r_3^{-1}\right) \right].
\eq
All harmonic polylogarithms 
can again be expressed in terms of $\overline{\mathrm{E}}$-functions or $\mathrm{ELi}$-functions.
For the harmonic polylogarithms of weight $1$ and $2$, the explicit expressions have been
given in eq.~(\ref{HPL_weight_1_2}),
for the harmonic polylogarithms of weight three 
the explicit expressions are given in appendix~\ref{appendix_harmonic_polylogs}.
We have verified numerically the results by comparing with the program {\verb|sector_de|}-{\verb|composition|} \cite{Bogner:2007cr}.


\section{Conclusions}
\label{sec:conclusions}

In this paper we have shown that
the Laurent series of the two-loop kite integral in $D=4-2\eps$ space-time dimensions
can be expressed in each order of the Laurent expansion
in terms of $\overline{\mathrm{E}}$-functions (or $\mathrm{ELi}$-functions).
We presented an algorithm to compute the term of order $\eps^j$ of the Laurent expansion.
As an example, we explicitly presented the first three terms of the Laurent expansion.
We expect that the class of $\mathrm{ELi}$-functions will be useful for a wider class of Feynman
integrals.
In fact, the $\eps^0$-term of the three-loop banana graph with equal masses in two space-time dimensions
has been recently computed in \cite{Bloch:2014qca} in terms of related functions.

\subsection*{Acknowledgements}

We would like to thank L. Tancredi for useful communication.
L.A. is grateful for financial support from the research training group GRK 1581.
C.B. thanks Deutsche Forschungsgemeinschaft for financial support under the project BO4500/1-1 and Humboldt University for hospitality.


\begin{appendix}

\section{Harmonic polylogarithms of weight three in terms of $\mathrm{ELi}$-functions}
\label{appendix_harmonic_polylogs}

The variables $y=t/m^2$ and $q=e^{i\pi\tau}$ are related by
\bq
 y & = & 
 - 9  
 \frac{\eta\left(\tau\right)^4 \eta\left(\frac{3\tau}{2}\right)^4 \eta\left(6\tau\right)^4}
      {\eta\left(\frac{\tau}{2}\right)^4 \eta\left(2\tau\right)^4 \eta\left(3\tau\right)^4},
\eq
where $\eta(\tau)$ denotes Dedekind's $\eta$-function
\bq
 \eta\left(\tau\right)
 & = &
 e^{\frac{\pi i \tau}{12}} \prod\limits_{n=1}^\infty \left( 1- e^{2 \pi i n \tau} \right)
 =
 q^{\frac{1}{12}} \prod\limits_{n=1}^\infty \left( 1 - q^{2n} \right).
\eq
Eq.~(\ref{integration_kernel_1}) and eq.~(\ref{integration_kernel_1a}) 
in combination with eq.~(\ref{Ebar_iteration}) and eq.~(\ref{Ebar_iteration2})
allow us to express any harmonic polylogarithm in the letters $\{0,1\}$ without trailing zeros 
in terms of $\overline{\mathrm{E}}$-functions.
At weight three we have
\bq
\lefteqn{
 G\left(0,0,1;y\right)
 =
 3 
 \left[ 
          \overline{\mathrm{E}}_{3;2}\left( -1;  1; -q \right)
        - \overline{\mathrm{E}}_{3;2}\left( r_6; 1; -q \right)
 \right]
 - 12
 \left[ 
          \overline{\mathrm{E}}_{0,1;-1,0;4}\left( r_3,-1;  -1,1; -q \right)
 \right.
 } & & \nonumber \\
 & &
 \left.
        - \overline{\mathrm{E}}_{0,1;-1,0;4}\left( r_3,r_6; -1,1; -q \right)
 \right]
 - 12
 \left[ 
          \overline{\mathrm{E}}_{0,2;-1,1;2}\left( r_3,-1;  -1,1; -q \right)
 \right. \nonumber \\
 & & \left.
        - \overline{\mathrm{E}}_{0,2;-1,1;2}\left( r_3,r_6; -1,1; -q \right)
 \right]
 + 48
 \left[ 
          \overline{\mathrm{E}}_{0,0,1;-1,-1,0;2,2}\left( r_3,r_3,-1;  -1,-1,1; -q \right)
 \right. \nonumber \\
 & & \left.
        - \overline{\mathrm{E}}_{0,0,1;-1,-1,0;2,2}\left( r_3,r_3,r_6; -1,-1,1; -q \right)
 \right],
 \nonumber \\
\lefteqn{
 G\left(0,1,1;y\right)
 =
 9
 \left[ 
          \overline{\mathrm{E}}_{0,1;-1,0;4}\left( -1,-1;   1,1; -q \right)
        - \overline{\mathrm{E}}_{0,1;-1,0;4}\left( -1,r_6;  1,1; -q \right)
 \right. } & & \nonumber \\
 & & \left.
        - \overline{\mathrm{E}}_{0,1;-1,0;4}\left( r_6,-1;  1,1; -q \right)
        + \overline{\mathrm{E}}_{0,1;-1,0;4}\left( r_6,r_6; 1,1; -q \right)
 \right]
 \nonumber \\
 & &
 - 36
 \left[ 
          \overline{\mathrm{E}}_{0,0,1;-1,-1,0;2,2}\left( r_3,-1,-1;   -1,1,1; -q \right)
        - \overline{\mathrm{E}}_{0,0,1;-1,-1,0;2,2}\left( r_3,-1,r_6;  -1,1,1; -q \right)
 \right. \nonumber \\
 & & \left.
        - \overline{\mathrm{E}}_{0,0,1;-1,-1,0;2,2}\left( r_3,r_6,-1;  -1,1,1; -q \right)
        + \overline{\mathrm{E}}_{0,0,1;-1,-1,0;2,2}\left( r_3,r_6,r_6; -1,1,1; -q \right)
 \right],
 \nonumber \\
\lefteqn{
 G\left(1,0,1;y\right)
 =
 9
 \left[ 
          \overline{\mathrm{E}}_{0,2;-1,1;2}\left( -1,-1;   1,1; -q \right)
        - \overline{\mathrm{E}}_{0,2;-1,1;2}\left( -1,r_6;  1,1; -q \right)
 \right. } \nonumber \\
 & & \left.
        - \overline{\mathrm{E}}_{0,2;-1,1;2}\left( r_6,-1;  1,1; -q \right)
        + \overline{\mathrm{E}}_{0,2;-1,1;2}\left( r_6,r_6; 1,1; -q \right)
 \right]
 \nonumber \\
 & &
 - 36
 \left[ 
          \overline{\mathrm{E}}_{0,0,1;-1,-1,0;2,2}\left( -1,r_3,-1;  1,-1,1; -q \right)
        - \overline{\mathrm{E}}_{0,0,1;-1,-1,0;2,2}\left( -1,r_3,r_6; 1,-1,1; -q \right)
 \right. \nonumber \\
 & & \left.
        - \overline{\mathrm{E}}_{0,0,1;-1,-1,0;2,2}\left( r_6,r_3,-1;  1,-1,1; -q \right)
        + \overline{\mathrm{E}}_{0,0,1;-1,-1,0;2,2}\left( r_6,r_3,r_6; 1,-1,1; -q \right)
 \right],
 \nonumber \\
\lefteqn{
 G\left(1,1,1;y\right)
 =
 27
 \left[ 
          \overline{\mathrm{E}}_{0,0,1;-1,-1,0;2,2}\left( -1,-1,-1;   1,1,1; -q \right)
 \right. } & & \nonumber \\
 & & \left.
        - \overline{\mathrm{E}}_{0,0,1;-1,-1,0;2,2}\left( -1,-1,r_6;  1,1,1; -q \right)
        - \overline{\mathrm{E}}_{0,0,1;-1,-1,0;2,2}\left( -1,r_6,-1;  1,1,1; -q \right)
 \right. \nonumber \\
 & & \left.
        + \overline{\mathrm{E}}_{0,0,1;-1,-1,0;2,2}\left( -1,r_6,r_6; 1,1,1; -q \right)
        - \overline{\mathrm{E}}_{0,0,1;-1,-1,0;2,2}\left( r_6,-1,-1;   1,1,1; -q \right)
 \right. \nonumber \\
 & & \left.
        + \overline{\mathrm{E}}_{0,0,1;-1,-1,0;2,2}\left( r_6,-1,r_6;  1,1,1; -q \right)
        + \overline{\mathrm{E}}_{0,0,1;-1,-1,0;2,2}\left( r_6,r_6,-1;  1,1,1; -q \right)
 \right. \nonumber \\
 & & \left.
        - \overline{\mathrm{E}}_{0,0,1;-1,-1,0;2,2}\left( r_6,r_6,r_6; 1,1,1; -q \right)
 \right].
\eq

\end{appendix}
\bibliography{/home/stefanw/notes/biblio}
\bibliographystyle{/home/stefanw/latex-style/h-physrev5}

\end{document}